\documentclass[preprints,article,accept,moreauthors,pdftex]{Definitions/mdpi}

\firstpage{1} 
\pubvolume{1}
\issuenum{1}
\articlenumber{0}
\pubyear{2021}
\copyrightyear{2020}
\datereceived{} 
\dateaccepted{} 
\datepublished{} 
\hreflink{https://doi.org/} 

\Title{Digging into Axion Physics with (Baby)IAXO}

\TitleCitation{Digging into Axion Physics with (Baby)IAXO}

\Author{Theopisti Dafni $^{1}$*\orcidA{} and Javier Galan $^{1}$*\orcidB{}}

\AuthorNames{Theopisti Dafni and Javier Galan}

\AuthorCitation{Dafni, T.; Galan, J.}

\address{
$^{1}$ \quad Centro de Astropartículas y Física de Altas Energías, (CAPA), Universidad de Zaragoza, C/ Pedro Cerbuna 12, 50009 Zaragoza, Spain}
\corres{Correspondence: tdafni@unizar.es (T.D.); javier.galan@unizar.es (J.G.)}

\abstract{Dark Matter searches have been ongoing for three decades; the lack of a positive discovery of the main candidate, the WIMP, after dedicated efforts, has put axions and axion-like-particles in the spotlight. The three main techniques employed to search for them complement each other well in covering a wide range in the parameter space defined by the axion decay constant and the axion mass. The International AXion Observatory (IAXO) is an international collaboration planning to build the fourth generation axion helioscope, with an unparalleled expected sensitivity and discovery potential. The distinguishing characteristic of IAXO is that it will feature an axion-specific magnet, with a large axion-sensitive cross-section, and will be equipped with x-ray focusing devices and detectors that have been developed for axion physics. In this paper, we review aspects that motivate IAXO and its prototype, BabyIAXO, in the axion and ALPs landscape. As part of this Special Issue, some emphasis is given on the Spanish participation in the project, of which CAPA is a strong promoter.}


\begin{document}
\section{Introduction}\label{sc:intro}

 The last century marked an era for the field of Particle Physics: not only did it offer the technological progress necessary to develop new detection techniques, it also brought the acceptance of special relativity and quantum mechanics as a basis to construct new physics models. These two circumstances guided the experimental physicists into a new paradigm: the Standard Model of particle physics (SM) was proposed, a frame in which, through quantum electrodynamics (QED) and quantum chromodynamics (QCD), the force interactions between the elementary particles could be described, and several of their characteristics predicted quite accurately.

The year 2012 proved to be a very important one: the Large Hadron Collider at CERN fulfilled its primary goal and discovered the Higgs boson. That was the last piece that researchers were looking for to complete the description of the SM. Today, the SM can explain with great accuracy all the fundamental interactions, excluding gravity. And yet, all that amounts to approximately 5\% of the matter in our universe: the rest is a dark mystery, with 68.5\% of the matter density of the Universe in the form of Dark Energy and a 26.5\% in the form of Dark Matter (DM).

Discovering that the universe has a dark side, has been a revolution to the way we see it and the way we try to understand it. Dark Energy being the most recent concept, there is still a lot of speculation regarding its nature; however, there is a lot more known about the particles that would constitute the DM. As predicted by the present cosmological model, the DM particles should have been produced in the early universe in copious amounts; they should be stable (with a decay time far exceeding the age of our universe), interacting weakly with ordinary matter, and have a mass.  The observational evidence for DM originates from gravitational interactions: the galaxy cluster velocities distribution at the Coma Cluster\,\cite{ZwickyComaCluster}, the rotation curves of galaxies\,\cite{VeraRubinRotation}, the Cosmic Microwave Background (CMB) anisotropies\,\cite{COBEAnisotropies}, and the gravitational lensing produced by massive objects in between the light path\,\cite{LensingDMDE}. They all require an additional invisible mass to describe the observations with the present understanding of Physics.

For more than 30 years now, experimentalists and theorists are trying to detect and identify the DM; a plethora of candidates has been put forward to solve the Dark Matter puzzle. Among them, the most sought after candidate is the Weakly Interacting Massive Particles (WIMPs), while in the recent years the axion is gaining weight in the terrain of direct searches. Both the axion and the WIMPs share the advantage of not having been hypothesized as a Dark Matter candidate; rather, they both appear in extensions of the SM in a natural and independent way. In principle, there is no reason to believe that the DM is composed only of one type of particles; the total observed DM mass could be produced by different contributing particles. Another typical candidate that could contribute to the Dark Matter pie are primordial black holes, that might have been produced in the early universe.

The axion emerged as the consequence of an elegant solution that was proposed to solve one of the puzzles that the SM cannot explain, the strong CP problem.  The interest in axions has generated a broad number of techniques to detect them, one of them being the helioscope. The aim of this article is to motivate and describe the planned baseline axion searches with the International AXion Observatory (IAXO), and give an outlook of its further potential.

The paper begins with a brief introduction to the axion theory, properties and motivation (Section~\ref{sc:motivation}); then follows a comment on the impact of the axion existence in cosmology and astrophysics, which constrain the axion properties (Section~\ref{sc:phenomenology}); this first part concludes with a context of axion detection techniques from a historical perspective (Section~\ref{sc:detection}). In the second part of this document we will present the initiative from the IAXO collaboration to build the next generation axion helioscope focusing on the different components of the instrument (Section~\ref{sc:IAXO}) and the Spanish leading role and contribution to the experiment (Section~\ref{sc:IAXOSpain}). Finally, we discuss the potential use of IAXO as a tool for exploiting different axion search techniques, serving as a future platform or infrastructure for maximizing the potential of discovery with new ideas or physics cases that could be part of  the existing infrastructure (Section~\ref{sc:observatory}).

\section{The Axion}\label{sc:motivation}

The axion is a hypothetical particle that emerged as a natural solution to the strong CP problem\,\cite{CHENG19881}. The strong interaction, described by the QCD theory, does not forbid a CP violation: the QCD Lagrangian admits a CP-odd term, thus allowing for CP violation in the strong interaction. Furthermore, since electroweak interactions violate CP, it is difficult to understand why nature has decided that strong interactions will not. The absence of CP-violation in strong interactions is quantified by the $\overline\theta$ parameter appearing at the corresponding CP-violating lagrangian term,

\begin{equation}
    \mathcal{L}_{\overline\theta} = -\overline\theta\frac{\alpha_s}{8\pi}G^a_{\mu\nu}\widetilde G^{a\mu\nu} 
\end{equation}

\noindent and it is measured experimentally to be unexpectedly close to zero, $|\overline\theta| < 1.8 \times 10^{-11}$, as it is deduced from the measurement of the electric dipole neutron moment,  $(0.0\pm1.1_{\mbox{stat}}\pm0.2_{\mbox{sys}})\times10^{-26}$\,e$\cdot$cm, its corresponding upper limit $d_n < 1.8 \times 10^{-26}$\,e$\cdot$cm\,\cite{PhysRevLett.124.081803}, and a recent calculation in QCD, $|d_n|<10^{-15}\cdot\overline\theta$\,e$\cdot$cm.\,\cite{PhysRevD.103.114507}. In principle, this phase could take any value between 0 and $2\pi$, since it is produced adding up QCD contributions of different nature, thus arising the question why this phase would have chosen a value such that the strong CP-violation term is canceled.

The Peccei-Quinn mechanism to restore CP conservation in the strong sector\,\cite{Peccei:1977hh,Peccei:1977ur} led soon into an interesting outcome - the axion - arising naturally as a dynamic solution to the fine tuning problem of the $\overline\theta$ parameter~\cite{Weinberg:1977ma,Wilczek:1977pj}, effectively, $\overline\theta$ is replaced by $\overline\theta + a /f_a$, where $a$ is the axion field, and $f_a$ is the energy scale at which the field symmetry would be spontaneously broken.

\subsection{Properties} 

The axion is a pseudo-scalar boson, and its properties are described by the theory through a unique parameter, the axion decay constant, or energy scale, $f_a$, which must be much greater than the electroweak scale in order to circumvent the constrains imposed by accelerator-based searches. The couplings of axions to ordinary matter and the axion mass, $m_a$, are both inversely proportional to $f_a$, leading to very weak couplings to ordinary matter.

Two main theoretical models are considered as a reference for QCD axion searches~\cite{DiLuzio:2020wdo}, hadronic axions or KSVZ axions~\cite{KimModified,Shifman:1979if}, and DSFZ or Grand Unification Theories (GUT) axions~\cite{Dine:1981rt,Zhitnitsky:1980tq}, which do not couple to hadrons at tree level. In these models both the couplings to ordinary matter and the axion mass, $m_a$, depend on $f_a$. Thus, we will find a relation between the different couplings to matter and $m_a$ that will constrain the regions we should look at to find QCD axions.

Furthermore, a plethora of axion-like particles (ALPs) with analogous properties emerge in different extensions to the SM. The main difference with the originally predicted axion resides in the lack of a specific relation between couplings and $f_a$, and therefore with its mass, $m_a$. Their similarity with standard QCD axions broadens the physics program of IAXO, permitting the exploration of ALPs-hinted regions wider than the theoretical QCD models~\cite{Irastorza:2018dyq,Graham:2015ouw}.

\subsection{Motivation}

The axion is strongly motivated already from a theoretical perspective. 
Its detection could be the key giving access to a new world of pseudo-scalar particles predicted by string theories.
Phenomenologically speaking, finding the axion is strongly welcome in physics since it could play a role on stellar evolution, cosmology, and propagation of radiation in the intergalactic medium, among the most relevant axion physics cases that will be detailed later in Section~\ref{sc:phenomenology}. Additionally, the axion could provide the key to unveil other poorly understood phenomenology, as for example, the Sun's interior magnetic fields\,\cite{PhysRevD.102.043019} or the solar corona problem\,\cite{Zioutas:2007xk,PhysRevD.98.103527,RUSOV2021100746}.
 
The axion properties and its interactions with ordinary matter, specially with light, make this particle very attractive from an instrumentation perspective. The coupling to photons has given rise to a number of experimental techniques exploiting the mastering of light in the form of laser experiments, resonant cavities, or quantum magnetic effects. These techniques give an idea of the versatility and diversity of experimental axion searches, awakening the creativity of physicists designing new and more challenging experimental setups.

The discovery of the axion would provide an additional tool to observe the universe, just as the discovery of neutrinos or gravitational waves have contributed to the new era of multi-messenger astronomy. The axion is the dream of physicists, from a theoretical, phenomenological and instrumental perspective.

\section{Axion phenomenology}\label{sc:phenomenology}

As a consequence of its very weak couplings to ordinary matter the axion results to be a long-lived particle, and therefore, under certain conditions, it is considered to be a potential dark matter candidate. Particularly appealing are the regions of the parameter space that address the DM and the QCD problem at once.
The parameter space of axions is strongly constrained by cosmological and astrophysical arguments, such as the evolution of stars or the expected cosmological axion abundance.

\subsection{Cosmological constrains}

The first studies including cosmological axion production through a mechanism called vacuum realignment\,\cite{PRESKILL1983127,ABBOTT1983133,DINE1983137} showed that the DM axion would overclose the universe for masses above $m_a\gtrsim10^{-6}$\,eV. This fact motivated the construction of axion haloscopes (see Section~\ref{sc:haloscopes}), searching for axions with properties that match those of a model with a dominant axion contribution to dark matter. However, the axion mass that would lead to DM axions (those axions solving at the same time the dark matter puzzle) seems to be hard to predict by the theory. Its value has been relaxed in the latest years by additional production mechanisms, such as axion string radiation, predicting that the mass of the DM axion could be at the meV scale\,\cite{PhysRevD.82.123508,10.21468/SciPostPhys.10.2.050}\footnote{It should be noted that recent calculations indicate a mass in the range of $\mu$eV~\cite{Buschmann:2021sdq}. The question of the mass-range is still open}. 
The next generation axion helioscopes (see Section~\ref{sc:helioscopes}) will provide an apparatus able to probe those regions. In this context, the search for the so-called \emph{ALP-miracle} is also particularly attractive, where an ALP at the meV region accounts for the dark matter of the universe and drives inflation at the same time\,\cite{ALPmiracle2017}. Scalar fields naturally appear as a way to explain a cosmological model with dark energy, such as quintessence fields\,\cite{PhysRevLett.75.2077}, chameleons\,\cite{PhysRevD.69.044026,PhysRevLett.93.171104} and other exotic candidates (see e.g.\,\cite{Irastorza:2018dyq} and references therein). The signatures produced by those exotic candidates share common features with the axion detection, and have already been exploited in a campaign to detect chamaleons with the CERN Axion Solar Telescope (CAST) helioscope\,\cite{ANASTASSOPOULOS2015172}.

\subsection{Astrophysical constrains}

\begin{figure}[H]
	\begin{center}
		\includegraphics[width=0.65\textwidth]{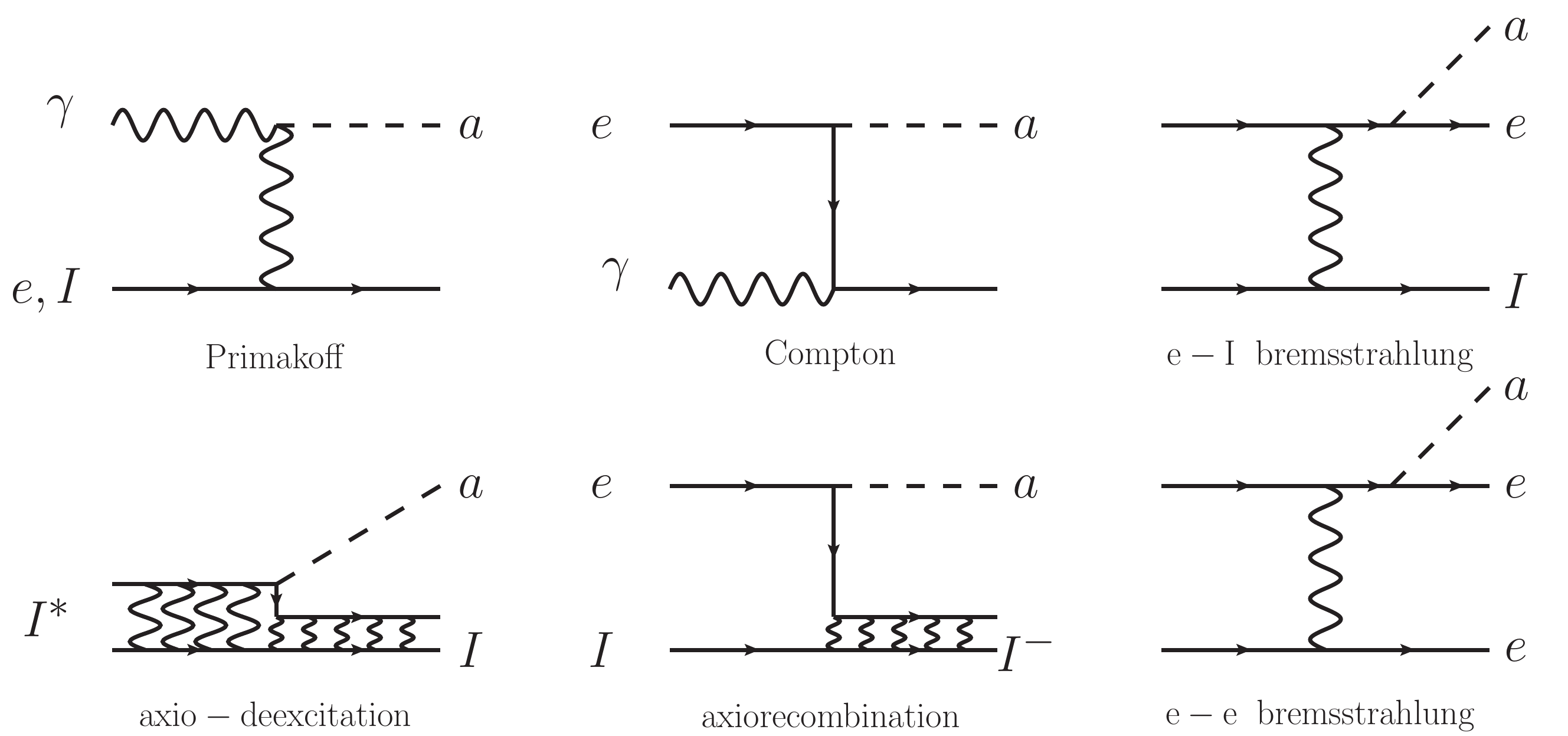} 
	\end{center}
	\caption{Feynman diagrams of the different processes responsible for axion production in the Sun. The Primakoff conversion of photons in the electromagnetic fields of the solar plasma is proportional to $g^2_{a\gamma}$, and it is present at almost any axion model. In non-hadronic models, such as DSFZ, non-negligible couplings to electrons open up new channels, such as Atomic axio-deexcitation and axio-recombination, axio-Bremsstrahlung in electron-ion or electron-electron collisions and Compton scattering with axion emission. Those additional channels are usually referred as the ABC solar axion flux, being the flux proportional to $g^2_{ae}$. Figure from ref~\cite{Redondo:2013wwa}.}
	\label{fig:processes}
\end{figure}

Observations from astrophysical origin produce even more stringent constrains\,\cite{Raffelt:2006cw, Giannotti:2015kwo, Giannotti:2017hny, DiLuzio:2021ysg}. A star is a rich particle physics laboratory where we find many different physics processes and interactions where axions could be produced inside the stellar medium\,\cite{Redondo:2013wwa} (see Figure\,\ref{fig:processes}). 

The existence of the axion, or ALPs, could play a role on the evolution of the star since, just as neutrinos, they would escape the sun generating an additional energy loss channel. Depending on the axion coupling strength, this additional energy loss channel would shorten the lifetime of the star. The age of our sun is well constrained by helioseismological observations and the measured neutrino flux, providing a first upper limit on the axion-photon coupling, $g_{a\gamma}\leq4.1\times10^{-10}$GeV$^{-1}$ (at 3$\sigma$)\,\cite{Vinyoles:2015aba}.

Studying the evolution of stars through a Hertzsprung-Russel (HR) diagram one finds an even more accurate measurement of the stars evolution by exploiting the statistics of stars at different evolutionary stages. The strongest bound comes from the Horizontal Branch (HB) to Red Giant Branch (RGB) ratio. The presence of an additional energy loss would reduce this ratio for a non zero $g_{a\gamma}$ value. This fact leads to an improved upper limit of $g_{a\gamma}<0.66\times10^{-10}$GeV$^{-1}$ (at 2$\sigma$)~\cite{Ayala:2014pea}.

The phenomenology in the skies is abundant; if axions are present, events such as supernovae may carry away an additional energy loss during the collapse, producing a non-negligible amount of axions by the extremely dense and hot medium. Different axion processes intervene in the physics of the supernovae, but the most dominant process is the nuclear bremsstrahlung, which exploits the properties of the measured neutrino burst of SN1987A\,\cite{PhysRevLett.58.1490} to obtain a limit on the axion-nucleon coupling, $g^2_{aN}\equiv g^2_{an}+0.61g^2_{ap}+0.53g_{an}g_{ap}\lesssim8.26\times10^{-19}$\,GeV$^{-2}$, recently revisited in the following reference\,\cite{Carenza_2019}.

\subsection{Astrophysical hints}
Furthermore, the measured value for the HB to RGB ratio seems to be lower than expected. Considering this effect could be coming from an energy loss through an axion Primakoff channel, we obtain what is known as the \emph{HB hint} value for $g_{a\gamma} = (0.29\pm0.18)\times10^{-10}$GeV$^{-1}$ (at 1$\sigma$)\,\cite{Ayala:2014pea, Straniero:2015nvc}.

The Primakoff emission is dominant at Sun-like stars, where the density of the hot plasma is low enough to avoid production suppression due to the effective photon mass in the medium. Stars at different evolution stages, such as white dwarfs (WD) or neutron stars (NS) require other axion cooling channels, as axion-electron coupling, $g_{ae}$, or axion-nucleon coupling, $g_{aN}$. Another strong constrain arises for non-hadronic axion models through the evolution of WDs. The WD luminosity function (WDLF) does not reproduce the theoretical expectation, however, introducing an axion in the range  $g_{ae}=1.12-4.48\times10^{-13}$ corresponding to the axion mass range $m_a$cos$(\beta)=4- 16$\,meV  (where cos$(\beta)$ is a free, model-dependent parameter that is usually set equal to unity) would reproduce the observational results\,\cite{10.1093/mnras/sty1162}. This result is consistent with the cooling observed through an independent measurement of the change of period of a single WD, which leads to $g_{ae}=4.8\times10^{-13}$\,GeV$^{-1}$\,\cite{WDCorsico2012} for a $m_a$cos$(\beta)=17.1$\,meV. The cooling rate of few available single WDs measurements are also coherent with the presence of an anomalous cooling mechanism\,\cite{CorsicoReview, Corsico:2019nmr}.

Another puzzle today is the lower attenuation of the extragalactic background light (EBL) than what expected: high energy photons travelling long distances in the universe should interact with the EBL producing positron-electron pairs, which would be translated into an effective cut-off on the gamma energy distribution observed on Earth. However, it seems that a good number of these high energy photons do reach the Earth. The oscillation of those photons into ALPs and back again to photons in the presence of intergalactic magnetic fields could enlarge the effective optical depth and explain this anomaly\,\cite{MirizziTHint2017}. Some authors identify this effect ---that we denote as the universe transparency hint, \emph{T-hint}--- with a low mass ALP of $m_a\sim7\times10^{-10}- 5\times10^{-8}$\,eV, and an axion photon coupling of $g_{a\gamma}\sim 1.5\times10^{-11}-8.8\times10^{-10}$\,GeV$^{-1}$\,\cite{PhysRevD.96.051701} to explain the excess observed by the CIBER collaboration\,\cite{CIBER2017}, while other authors constrain the region nearby at the $g_{a\gamma}\sim10^{-12}-10^{-11}$GeV$^{-1}$ level, for axion masses in the range $m_a\sim1-100$\,neV using HESS measurements\,\cite{BRUN201725} and Fermi-LAT observations\,\cite{CHENG2021136611}.

\section{Detecting axions and ALPs}\label{sc:detection}

Many models have been formulated to describe the possible interactions of axions and ALPs with matter and fields, from which a combination of couplings to nucleons, electrons and photons (including the processes previously shown in Figure~\ref{fig:processes}) are integrated to describe a complete theoretical framework. This theoretical description provides guidance to instrumentalists on the construction of experimental apparatuses that aim at the detection of such an elusive particle, and tune the sensitivity of a particular setup to explore a theoretically preferred region of the axion parameter space.

\begin{figure}[h]
\begin{center}
\includegraphics[height=8.5cm]{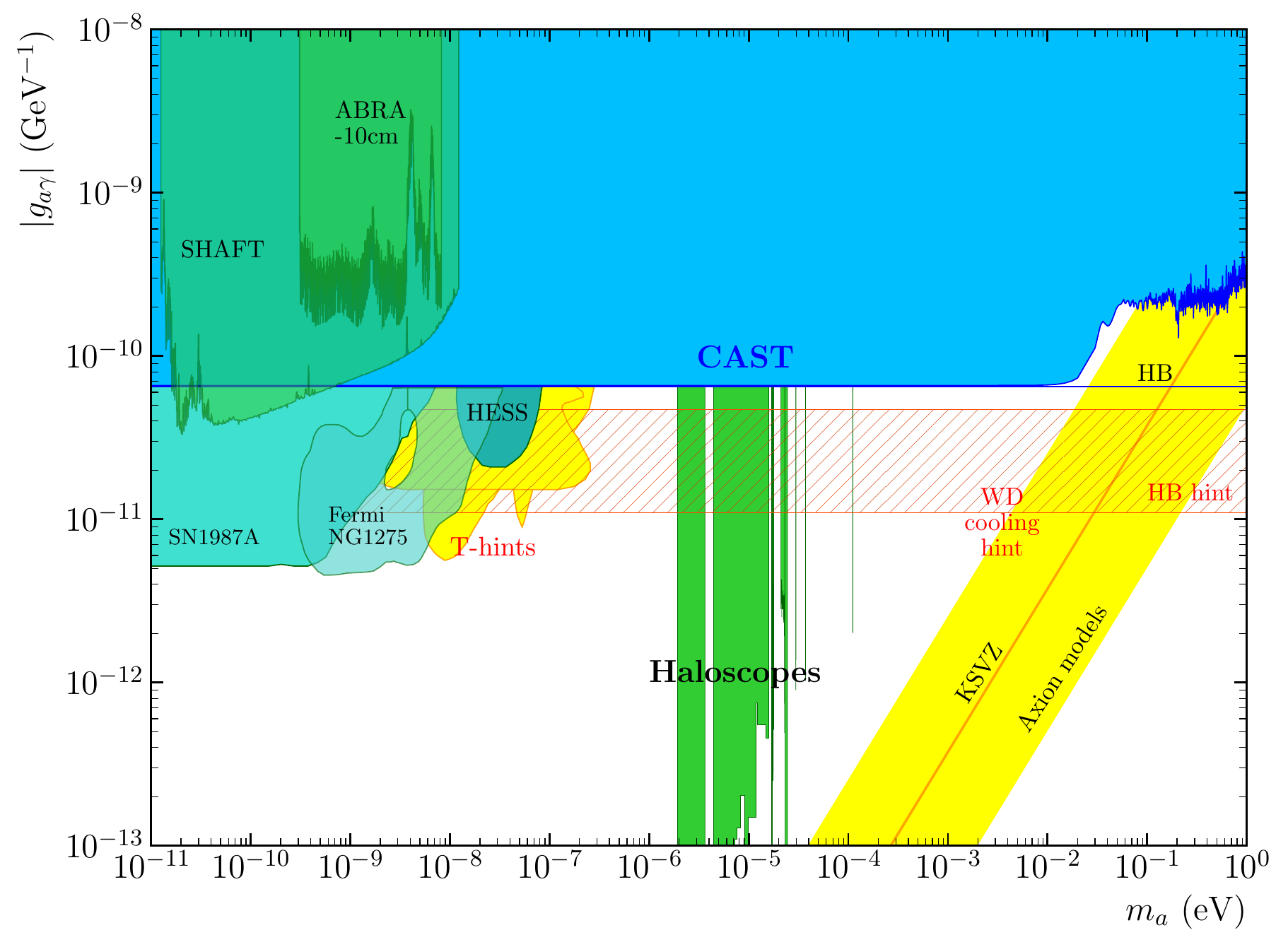}
\caption{\label{fig:exclusionRaw} The present axion searches experimental landscape showing the reached sensitivity in terms of the axion-photon coupling, $g_{a\gamma}$, for different experimental searches (CAST\,\cite{Anastassopoulos:2017ftl}, Haloscopes\,\cite{ADMX:2009iij}, SHAFT\,\cite{Gramolin:2020ict}, ABRA-10cm\,\cite{Salemi:2021gck}). Together with the different astrophysical constrains (SN1987A, HB, HESS, Fermi-LAT) and astrophysical hints (T-hints, WD cooling hints, and HB hint) already described in section~\ref{sc:phenomenology}. The yellow band denotes the region of the parameter space favored by QCD axion models.}
\end{center}
\end{figure}

In particular, the coupling relations constrained by the KSVZ and DSFZ models define the regions of the $g_{a\gamma}-m_a$ parameter space favored by QCD axions\,\cite{DiLuzio:2020wdo}. Figure~\ref{fig:exclusionRaw} shows those regions together with the astrophysical hints (described previously in Section~\ref{sc:phenomenology}), and the experimental limits that will be later described in this section. 

Originally, it was proposed that $f_a$ could be of the same order as the electroweak scale; this implied an axion with rather strong couplings and a mass of the order of 100$\;$keV and was soon excluded by experimental data (for example beam dump experiments\,\cite{Bellotti:1978hg,Ellis:1978my,Bechis:1979kp,Faissner:1980qg}). As already discussed (Section~\ref{sc:phenomenology}), considerations of the stellar evolution contribute also to limit the axion mass to lower values, being the stringent globular cluster limits valid up to about $\sim$100\,keV\,\cite{Carenza:2020zil}.

If the scale were to be much higher, of the order of 10$^{15}\;$GeV, considered in GUT, then the axion mass would be very small, $m_a\sim10^{-9}$\,eV. Such an axion would actually evade detection, in the type of experiments of the time, and was dubbed \emph{invisible}. 
Then, in 1983, Pierre Sikivie showed how there were ways of detecting the axion~\cite{PhysRevLett.51.1415}; he proposed to use \emph{axion haloscopes} in order to probe axions of cosmological origin, and \emph{axion helioscopes} for the detection of axions emitted by the Sun. 
In the recent years, several experimental techniques have emerged that exploit the interactions of the axions and ALPs, and which are sensitive to certain couplings or products of these: an extensive review of most of these techniques is given in~\cite{Irastorza:2018dyq,RevModPhys.93.015004}.

Here we will briefly mention the three main categories that are sensitive to the axion-photon coupling, the most generic coupling of axions and ALPs, present in most of the models. This separation is based on the origin of the axions: the aforementioned haloscopes and helioscopes, and pure laboratory experiments. 

\subsection{Pure laboratory experiments}
The pure laboratory experiments, or \emph{photon regeneration experiments}, include mainly two types of experiments, in which the aim is to produce and immediately detect the axions with the help of a photon beam (laser) that passes through strong magnetic fields. 

One type, also known as \emph{light-shining-through-wall} (LSW), requires two consecutive areas with a strong magnetic field. In the first, delimited by an optical wall, a laser is travelling back and forth perpendicular to the field and axions are expected to be born; the wall marking the second one, will allow only the axions to pass through, and that is where they are expected to reconvert into detectable photons of energies similar to those of the injected laser. Several experiments of this type have been proposed and performed: the ALPS-II project\,\cite{Bahre:2013ywa}, under construction at DESY, has the highest expected sensitivity of such an experiment to date, approaching the limit given by astrophysics. 
In the second, experiments that are exploring the properties of QED using polarimeters to measure magnetic birefringence and dichroism, can be sensitive to axions and ALPs: laser photons are converted into axions inside the polarimeter; the photons resulting from the re-conversion of axions will have a different polarization from the injected light (laser) due to the effect of the electric field parallel to their propagation axis. PVLAS~\cite{PVLAS:2005sku, PVLAS:2007wzd} was one of the experiments that searched for axions in their setup; the announcement of a possible positive signal, which was later retracted, gave birth to a large amount of theoretical, phenomenological and experimental papers and a commotion that promoted the axion and ALP field enormously . 

Despite the fact that the sensitivity of these experiments for the moment does not reach that of the haloscopes and the helioscopes, the great advantage they hold against the others is that they are their own source of axions, depending much less on theoretical models and parameters. 

\subsection{Haloscopes}\label{sc:haloscopes}

Haloscopes do not need to produce their own axions, as they are conceived to detect the axions of cosmological origin that populate the local Dark Matter halo. 

The principle of detection here is based on the assumption that axions could convert into detectable photons inside a resonant microwave cavity permeated by a magnetic field. The sensitivity achieved with this technique is very high, however only to the mass corresponding to the cavity resonance frequency. Therefore, it is convenient that this frequency be adjustable, in order to allow for the scanning in several axion masses. ADMX~\cite{ADMX:2009iij, ASZTALOS201139} has been the leading experiment for several decades and has reached the highest sensitivity in the $g_{a\gamma}$ for several axion masses of the order of $10^{-6}$eV, covering a major region of the haloscope searches in Figure~\ref{fig:exclusionRaw}. A good number of new ideas have been proposed in the recent years either with cavities, dielectric haloscopes and even very small changes in magnetic fields, all of which nicely complement the ADMX experiment\,\cite{RevModPhys.75.777,Melc_n_2018,CAST:2020rlf}. However, one drawback of haloscope searches is that they are very dependent on the DM halo models employed, and their sensitivity relies on the axion being a dominant component of the DM.

\subsection{Helioscopes}\label{sc:helioscopes}
In contrast to DM axions, whose density is fixed and exist since very early in the universe, one can have a copious amount of \emph{fresh} axions, readily produced in the core of stars, where any number of the possible interactions shown in Figure~\ref{fig:processes} could take place. If the star in question is the Sun, then one talks about helioscopes.

\begin{figure}[t]
	\begin{center}
		\includegraphics[width=0.6\textwidth]{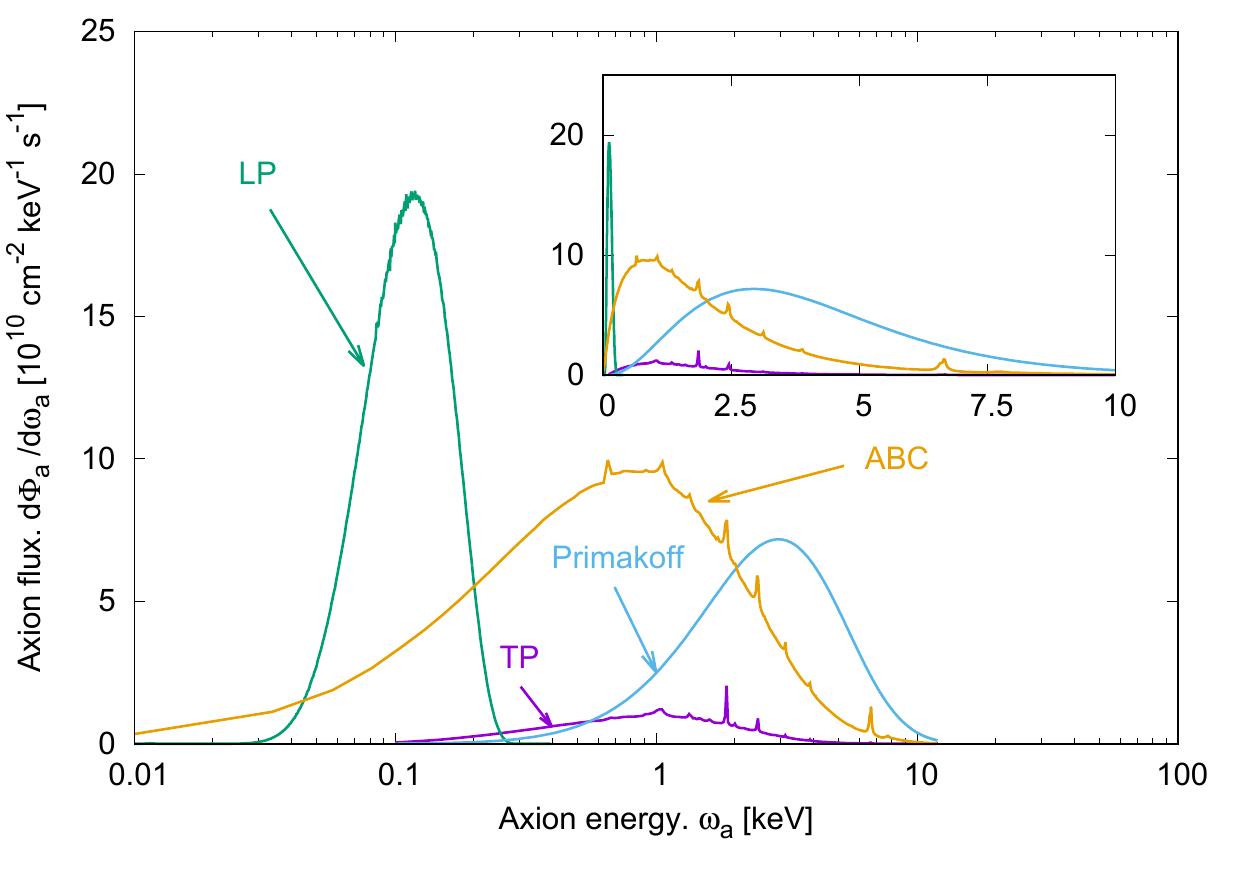}
	\end{center}
	\caption{The different solar axion components produced in the interior of the Sun. The recently calculated longitudinal plasmon (LP) and transversal plasmon (TP)~\cite{PhysRevD.101.123004,PhysRevD.102.123024}, together with the more conventional and well known axion flux components; Primakoff and electron coupling induced flux through Atomic, Bremsstrahlung and Compton axion-electron interactions (ABC)\,\cite{PhysRevD.33.897,Redondo_2013}. A linear energy scale is shown in the embedded plot.}
	\label{fig:solar_flux}
\end{figure}

This technique, like the haloscopes, relies on the assumption that a flux of axions is already present. The dense and hot plasma conditions in the Sun allow the production of an amount of axions large enough to be detectable at Earth, so much that the sensitivity reached is much higher than in the laboratory experiments. The generated solar axion flux is made of various continuous and monochromatic components related with different production mechanisms, and they typically produce axions at the energies of x-rays (see Figure~\ref{fig:solar_flux}).

This flux of axions could be converted into detectable photons if one facilitates the conditions for the inverse Primakoff effect to take place. 
That could happen in the lattice of crystals; interestingly, there could be an enhanced effect depending on the incident angle and whether the Bragg condition for a coherent diffraction is met~\cite{Paschos:1993yf}. In this technique, a time-dependent signal is expected, as the incident direction with respect to the crystalline planes changes due to the movement of the Earth during the day and the year. A number of experiments, mainly focused on WIMP search, have looked for such a signal~\cite{SOLAX:1997lpz, COSME:2001jci, Belli:2012zz, Armengaud:2013rta}.

However, the most competitive helioscopes are those using an intense transverse magnetic field in order to convert the solar axions into photons, then use photon detectors operating after the magnet conversion volume to identify an excess while the magnet is aligned with the Sun. In the next section we will focus on this particular concept of helioscope, contextualizing its evolution in the last decades, and providing the basic recipes that play a role in the construction of the magnet-based helioscope technique originally proposed by Sikivie\,\cite{PhysRevLett.51.1415} and refined later on, in reference\,\cite{PhysRevD.39.2089}.

\subsubsection{Helioscope essentials in a nutshell}
An indication of the sensitivity potential of a helioscope can be given by the number of expected photons, $N_\gamma$, at the end of its magnet. $N_\gamma$ depends on 
the solar axion flux at the earth $\frac{{\rm d}\Phi_{\rm a}}{{\rm d}E}$, the observation time $t$, the axion-sensitive area of the magnet $A$ and the conversion probability of axions to photons, $P_{{\rm a}\to\gamma}$, integrated to the energy range where axions are expected:  

\begin{equation}\label{eq:nphotons}
 N_{\gamma} = \int \frac{{\rm d}\Phi_{\rm a}}{{\rm d}E}\,P_{{\rm a}\to\gamma}\,A\,{\it t}\,{\rm d}E \;.
\end{equation}

The conversion probability takes into account the momentum transfer between the incoming axion and the out-coming photon. The maximum conversion probability is reached when the momentum transfer is minimum, that is to say, when the axion and the photon are in phase. The axion mass-range to which a helioscope will be sensitive, depends on the magnetic field length $L$ and strength $B$. When the axion-sensitive volume is in vacuum, then there is an $m_a$ up to which $N_\gamma$ is maximum. If the axion has a higher mass, then the coherence condition will not be met within the specific magnet length and the sensitivity will rapidly drop. In order to restore the sensitivity, a refractive gas could be used: in this way, the photon will gain an effective mass and the mismatch will be compensated\,\cite{PhysRevD.39.2089}. For a maximum effect, the buffer-gas pressure (density) should be as close to $m_a$ as possible and sweeping the axion mass axis requires several data taking points in different pressures. Low-Z gases like $^4$He and $^3$He are the usual buffer gases employed in order to minimize the photon re-absorption in the gas.

The Rochester-Brookhaven-Fermi collaboration was the first to run a helioscope in the early nineties\,\cite{Lazarus:1992ry}. A 1.8$\,$m-long magnet reaching 2.2\,T was siting on a table oriented towards the sunset and taking data for 15\,min every day, using an x-ray proportional chamber.  
Later on followed SUMICO~\cite{Moriyama:1998kd, Inoue:2002qy, Inoue:2008zp}: the important step forward in this second generation helioscope was that the 4$\,$T, 2.3$\,$m-long magnet was sitting on a platform that could follow the Sun for 12h a day. SUMICO used PIN photodiodes as x-ray detectors and took data both under vacuum (for a week) and with helium gas inside the bores, imposing an upper limit on the axion-to-photon coupling for masses 0.84\,$\leq m_a\leq$\,1\,eV.  
In 1999, a third generation helioscope was proposed, the CERN Axion Solar Telescope (CAST). 

CAST~\cite{Zioutas:2004hi, Andriamonje:2007ew, Arik:2008mq, Aune:2011rx, Arik:2013nya, Arik:2015cjv, Anastassopoulos:2017ftl} has provided the most stringent upper limits on $g_{a\gamma}$ for axion masses of up to about 1\,eV, shown in Figure~\ref{fig:exclusionRaw}: for masses below $m_a$\,<\,0.2\,eV, the upper limit was set to $g_{a\gamma}$\,<\,0.66$\times10^{-10}$\,GeV$^{-1}$.
Like the previous searches, CAST took data with the magnet bores under vacuum and using buffer gases to extent the sensitivity to higher masses and for a broader axion mass range, for which the average upper limit was $g_{a\gamma}$\,<\,3.3$\times10^{-10}$\,GeV$^{-1}$, the exact value depending on each pressure setting.
The significant improvement that CAST achieved was thanks to an enhanced magnet, x-ray focusing devices and the excellent performance of its detector systems in terms of background reduction. As a magnet, CAST employed a decommissioned LHC prototype magnet that would reach 9\,T along its 9.2\,m, able to follow the Sun for 3\,h in total during sunrise and sunset. Equally important proved to be the main detector technologies employed in the experiment; microbulk Micromegas detectors which reached very low background levels\,\cite{TOMAS2012478}, on the one hand, and on the other, an X-ray focusing device from the XMM-Newton mission that was coupled to a CCD camera. The use of an x-ray telescope in CAST was pioneering and the distinguishing point from the previous helioscopes with regard to the discovery potential that it supposed. 
CAST also looked into the axion-electron coupling~\cite{Barth:2013sma} and several other possibilities on the axion- and ALP-physics frontier, some with data taken in a parasitic way to the mainstream solar axion search~\cite{Anastassopoulos2015, Andriamonje:2009dx, Andriamonje:2009ar, Anastassopoulos:2018kcs, ArguedasCuendis:2019fxj}.
The signal to noise ratio, that is, the number of signal photons $N_\gamma$ expected over the number of background photons $N_b$, defines the discovery potential of the experiment (and in absence of signal, gives the limit of its sensitivity). A rather instructive way of expressing it, with respect to the main aspects that take part in the axion-to-photon conversion and the subsequent detection of the photons, is the following: 

\begin{equation}
\dfrac{N_\gamma}{\sqrt{N_b}}\sim B^2 L^2 A \; \epsilon_d b^{-1/2} \; \epsilon_o a^{-1/2} \; \epsilon_t^{1/2} t^{1/2}\;, 
\label{eq:sens}
\end{equation}

\noindent with the total time of data-taking of the experiment $t$ and $\epsilon_t$, the fraction of time the magnet tracks the Sun; the length $L$ and the strength $B$ of the provided magnetic field as well as the axion-sensitive area $A$; the background level $b$ and their detection efficiencies $\epsilon_d$; and the efficiency $\epsilon_o$ and total focusing area $a$ of x-ray focusing optics.

Each one of the helioscope generations has been building up on the strengths of the previous one; however, they all recycled equipment from other experiments and put them to use in the axion quest, with CAST achieving a sensitivity that matches astrophysical constraints. With this legacy, the helioscope technique can be considered mature enough to stand on its own and demand dedicated infrastructure specifically designed for its purpose: finding the axion; and that is the goal of the International Axion Observatory project.

\section{IAXO: the new generation helioscope}\label{sc:IAXO}

\begin{figure}[H]
 \includegraphics[width=7.2 cm]{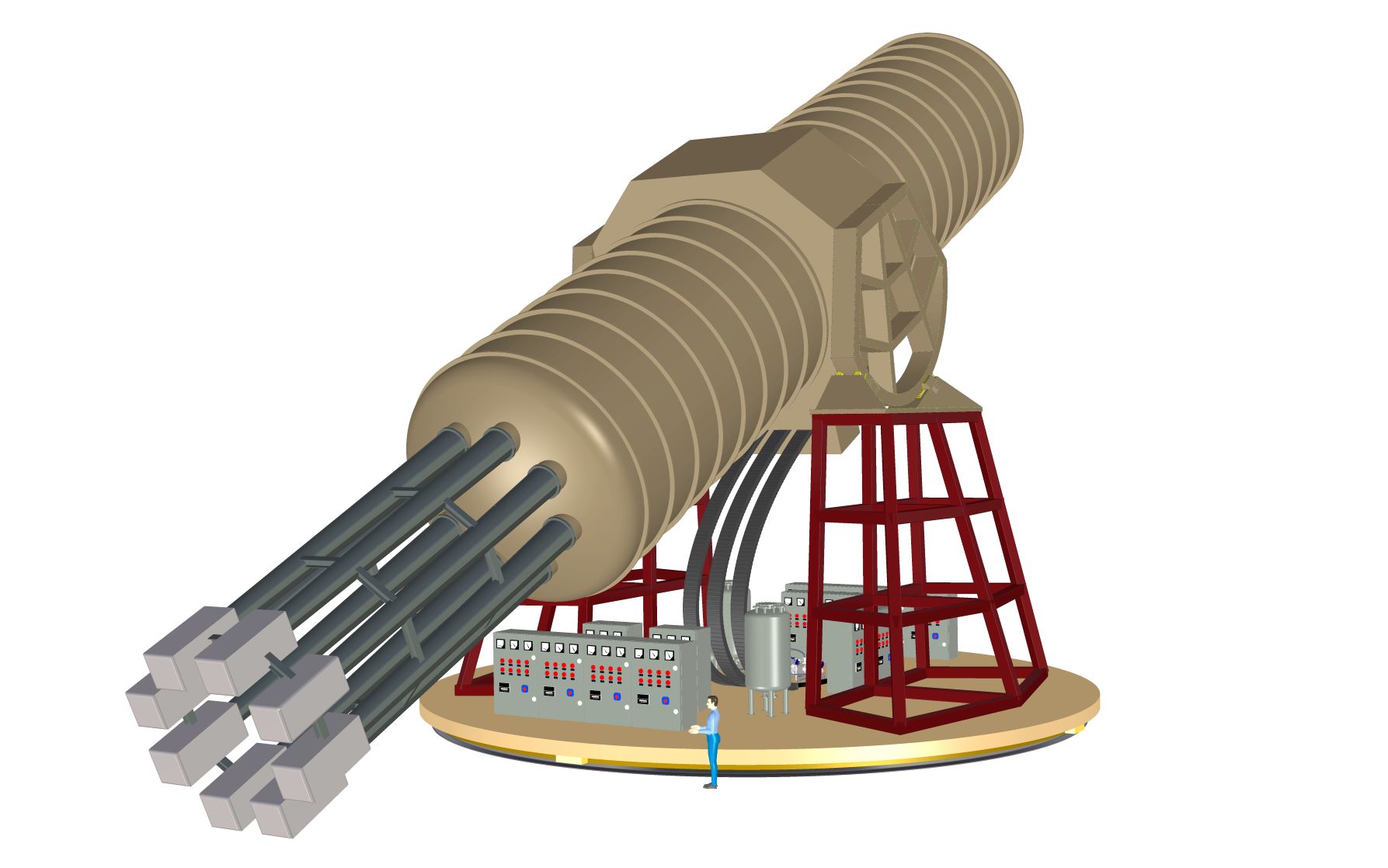}
 \includegraphics[width=5.2 cm]{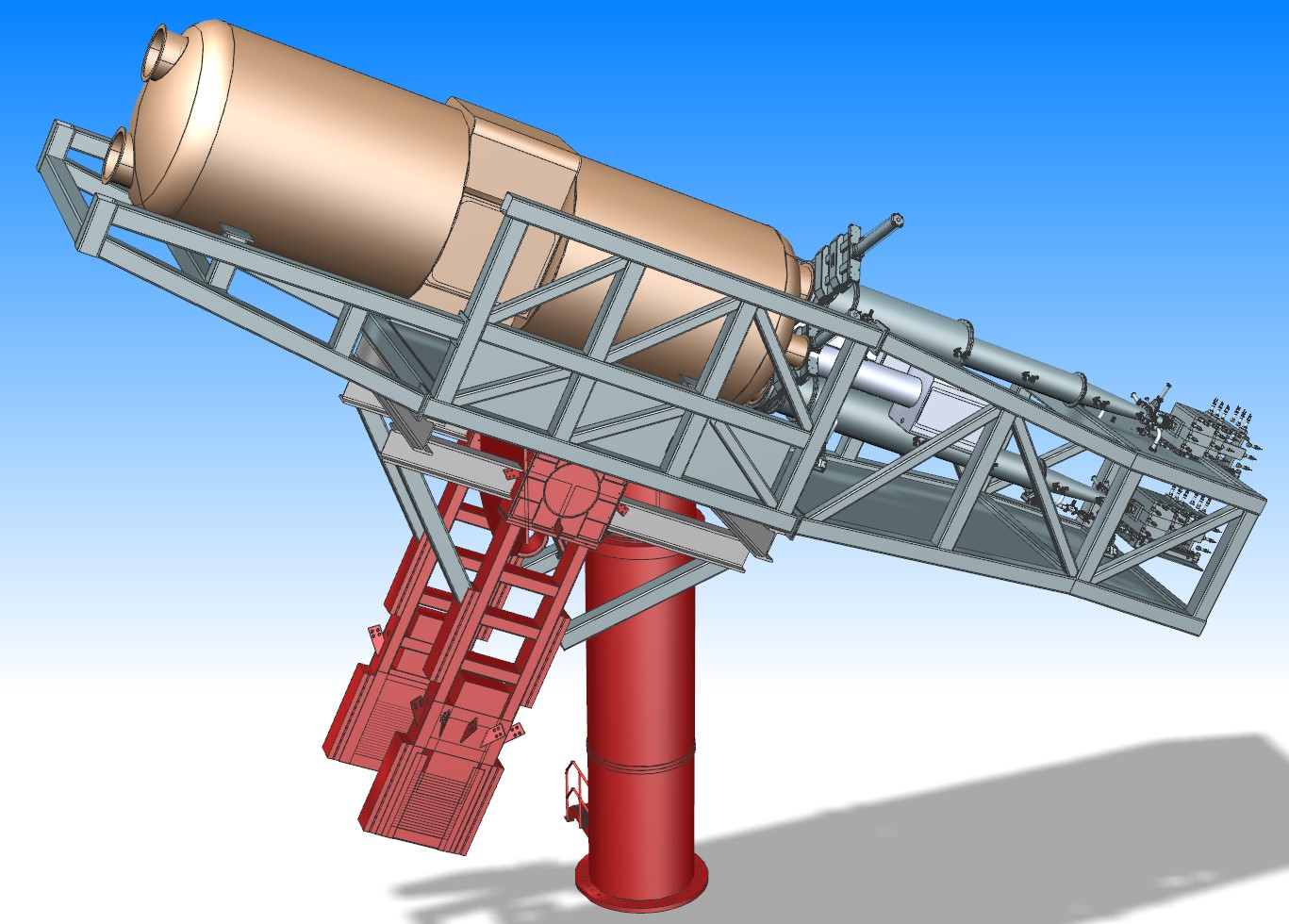}
\caption{(\textbf{a}) Sketch of IAXO: the 20\,m-long magnet features 8 bores with a 60\,cm diameter. Each bore will be equipped with an x-ray focusing device and low-background detectors, which, as indicated, will have an appropriate shielding. The moving platform will permit following the centre of the Sun for 12~h a day. (\textbf{b}) The 10\,m-long BabyIAXO magnet will also be tracking the Sun for approx. 12\,h a day. It will have two 70\,cm diametre bores, on which two different telescopes will be coupled.\label{fig:sketch}}
\end{figure}

IAXO is an international collaboration which integrates experts on various fields such as axion physics and phenomenology, magnets, x-ray focusing devices, detectors, low-background techniques, software and engineering~\cite{IAXO:2019mpb}. The collaboration is coordinating efforts to push further the axion physics frontier, maximizing the parameters of the sensitivity (Equation~\ref{eq:sens}), with the following characteristics~\cite{Irastorza:2011gs}:

\paragraph{Magnet} The IAXO magnet will have a toroidal design, similar to that of ATLAS; it will be 20\,m long and will feature 8 bores. Although the peak field will be of the order of 2.5\,T, which seems modest with respect to that of CAST, the aperture of each bore will be such that the total axion-sensitive area will amount to 2.3\,m$^2$, to be compared to 29\,cm$^2$ aperture in CAST.  Figure~\ref{fig:sketch} shows a sketch of the experiment, where the size of the magnet can be appreciated.
\paragraph{Moving platform} The magnet should be able to point to and follow the centre of the Sun with great precision, approximately one order of magnitude higher than the typical astronomical telescope. Another enhancement with respect to its predecessor is the exposure time; it is foreseen that the IAXO magnet will be able to follow the Sun for a minimum of 12\,h a day.
\paragraph{X-ray focusing devices} A focusing device concentrates all of the photons coming in from the bore aperture onto a very small area (spot). This offers a major advantage both in regard to the discovery potential of a helioscope and to technical aspects. In the first case, if axions are detected, then an intense signal is expected to appear in the spot while in the rest of the active area of the detector a simultaneous measurement of the background is performed; in the second, it allows for the use of small area detectors, which are generally easier to operate and shield. The main approach for IAXO to image the 3-arcmin of the centre of the Sun are segmented, slumped-glass optics like the one employed for the NUSTAR satellite~\cite{Harrison:2013md}. Each of the 8 bores of IAXO will be instrumented with an x-ray focusing device that will focus the x-ray photons from the converted axions onto a small spot of a low-background detector.
\paragraph{Low-background detectors} The requirement on the background levels of the detectors for IAXO is rather strict: it should be better than $10^{-7}$\,keV$^{-1}$\,cm$^{-2}$\,s$^{-1}$. The baseline option for the detector technology is small Time Projection Chambers with a microbulk Micromegas readout~\cite{Andriamonje:2010zz}. Their design is based on the IAXO pathfinder system which has been taking data in CAST in the last years with an unprecedented sensitivity, with background levels of $10^{-6}$\,keV$^{-1}$\,cm$^{-2}$\,s$^{-1}$~\cite{Aznar:2015iia}. 

In parallel, within the collaboration, there is an active R\&D for other technologies, like GridPix\,\cite{Krieger:2018nit, Anastassopoulos:2018kcs}, Metallic Magnetic Calorimeters (MMC)\,\cite{Unger_2021}, Neutron Transmutation Doped sensors (NTD), Transition Edge Sensors (TES) and Silicon Drift Detectors (SDD)\,\cite{Mertens_2015} which are being developed for their installation in IAXO in a later stage. 
Developing different detector technologies fulfilling the specifications of IAXO detectors is an advantage for the project in two senses. On one hand, a healthy competition and parallel detector analysis guarantees a more robust and independent result. And on the other hand, different technologies might be suited to different axion physics cases or components, such as measuring axions with very low energies (detectors with low energy threshold required) or measuring monochromatic energies with high precision (detectors with an excellent energy resolution required).
Low-background techniques are being applied in order to keep the levels to the desired order, that consist both on enclosing the detectors inside an appropriate shielding, and on screening campaigns aiming at reducing the intrinsic radioactivity levels of the detector systems.

\paragraph{}
One of the advantages of the IAXO project is that it is possible to use known technologies for all major systems, reducing the risk associated to the construction, but at the same time parallel R\&D programs in a long term planning are being addressed. In order to encourage research groups to define robust strategies for the development of appropriate technologies for IAXO and to assure that all the working groups are ready towards a major challenge, the collaboration decided to proceed in stages towards the completion of the physics program. The first one is the construction and operation of a scaled IAXO magnet capable of exploring new regions of the parameter space, and therefore with physics discovery potential itself, BabyIAXO.

\subsection{BabyIAXO}

BabyIAXO is conceived as a small-scale IAXO: it will consist of two, 10\,m-long magnet coils that will offer two bores of a diameter of 70\,cm each~\cite{Bykovskiy:2021oqs}. The x-ray optics and detectors to be used will have dimensions similar to the ones expected for IAXO. However, in addition to the custom-made optics, BabyIAXO will count with a spare telescope of the XMM Newton for the second bore. 

The requirements on pointing accuracy, in conjunction with BabyIAXO's size and weight, are similar to those used in the big scale gamma-ray experiments such as the Cherenkov Telescope Array (CTA). The moving platform of BabyIAXO is inspired by CTA's Medium-Sized Telescope (MST)~\cite{Garczarczyk:2015zya}.
A sketch can be seen on the right part of Figure~\ref{fig:sketch} and further details on the conceptual design of BabyIAXO can be found in~\cite{IAXO:2020wwp}.

As a technological prototype for IAXO, BabyIAXO will serve as a proof of concept of the technologies chosen for its construction, and it will help to consolidate the different groups of expertise in the design and construction of the apparatus. It is expected to pave the way to a smooth implementation of the IAXO systems, with the risk mitigation and technology preparation that this task encompasses. The project is amid final definitions and construction of the subsystems and is expected to start commissioning in 2024, hosted at DESY, most probably in the HERA South Hall. The construction and commissioning of IAXO will take place in parallel to this first physics runs of BabyIAXO.
Yet, one should not dismiss the importance of BabyIAXO's physics outcomes: BabyIAXO will produce relevant physics results, as it can be seen from the expected sensitivity discussed in the following.

\subsection{Expected performance} 
\begin{figure}[H]
\begin{center}
\includegraphics[width=12.5 cm]{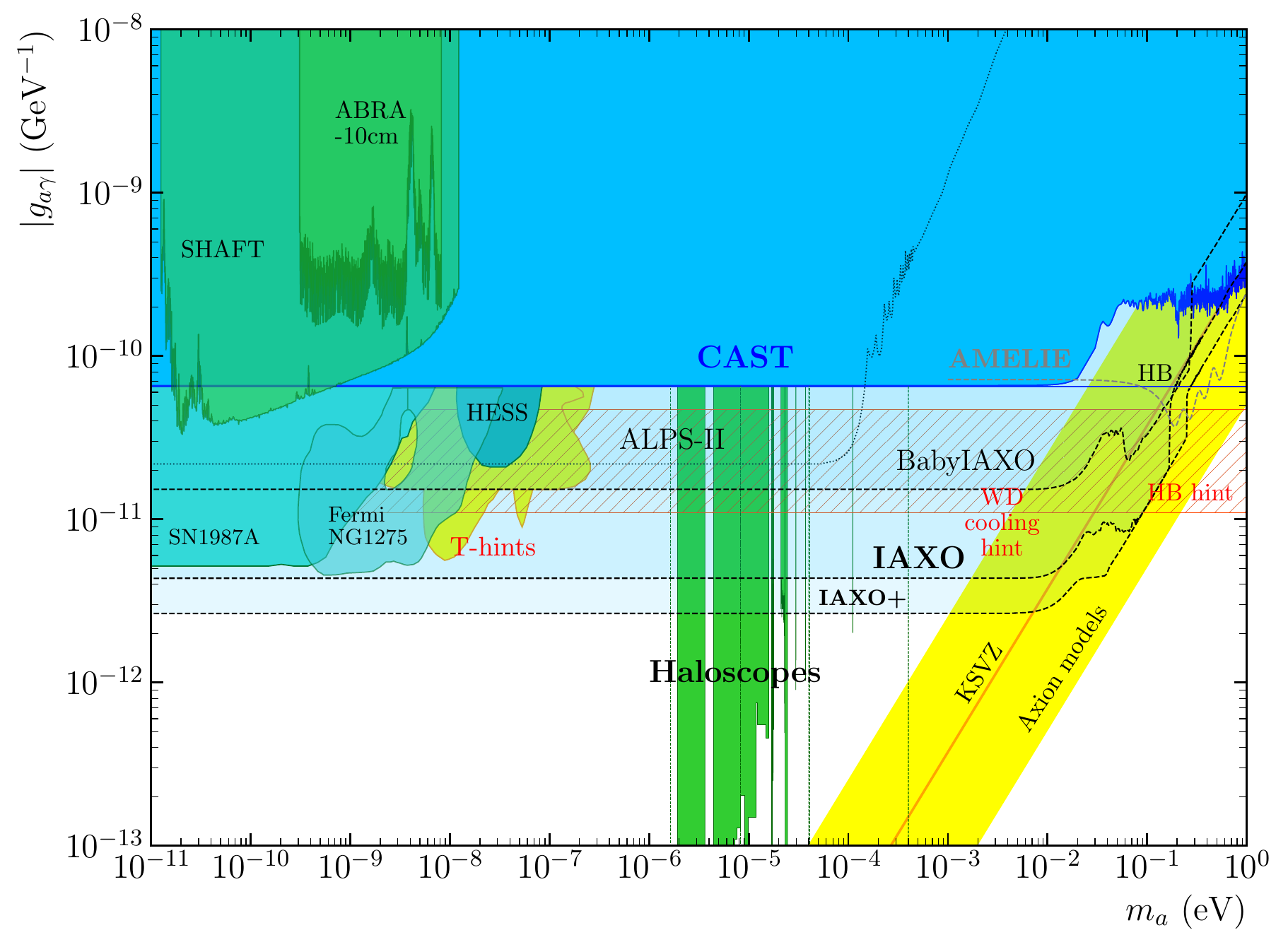}
\caption{ A plot of $g_a\gamma$ vs $m_a$, with the most stringent results (solid) and sensitivity prospects  (dashed) of observations and experiments, directly comparable with the previous Figure~\ref{fig:exclusionRaw}. Prospects of different phases of IAXO are shown, BabyIAXO, IAXO, and an upgraded version of IAXO, IAXO+. Other experimental prospects with potential to explore new regions of the parameter space are also included, as ALPS-II\,\cite{Bahre:2013ywa} or AMELIE\,\cite{Gal_n_2015}.
\label{fig:exclusion}}
\end{center}
\end{figure}

As stated in the beginning of the section, Equation~\ref{eq:sens} is used as a guide for the specifications and design of the main subsystems of (Baby)IAXO, defining suitable figures of merit for all systems involved and with the aim to maximize the parameters therein. In this way, the $N_\gamma/\sqrt{N_b}$ of IAXO will be over a factor 10$^4$ higher than that of CAST, which translates to one order of magnitude in terms of the coupling constant $g_{a\gamma}$. Figure~\ref{fig:exclusion} shows the $g_{a\gamma}$ vs $m_a$ parameter space, where the most relevant experimental results are drawn, along with the prospects of IAXO. It is clear that IAXO will cover a good part of the space, entering well into the QCD axion models band already in the vacuum phase, for $m_a<\rm10^{-2}$\,eV. For larger masses, up to approximately 1\,eV, IAXO will count with a data-taking phase using a buffer gas. 

In the same plot, one can find the projected sensitivity of BabyIAXO; it can be seen that BabyIAXO will have enough sensitivity to look into unexplored parts of the parameter space, as it is foreseen to surpass the $g_{a\gamma}^4$ of CAST. These projections are computed taking into account two data-taking campaigns, in vacuum and with a buffer gas, and under the assumption of absence of signal. 

BabyIAXO will allow to start probing regions of the parameter space that are strongly motivated by observations, approaching the \emph{HB-hint} and \emph{T-hint}, and covering partially the \emph{WD cooling anomaly hint}. At the same time, it will cover regions of the parameter space at the meV scale, favored by QCD axion models and/or connected with other fundamental physics problems, such as the DM or inflation. See reference\,\cite{DiLuzio:2021ysg} for a recent review on BabyIAXO sensitivity to those favored regions.

The discovery potential of BabyIAXO and IAXO is high as both will enter areas not probed so far. Should either of the two detect the axion, then important information could be extracted regarding its mass and coupling constants. Provided sufficient statistics is obtained, IAXO could measure the axion mass and the coupling to photons and electrons in an independent way~\cite{Dafni:2018tvj, Jaeckel:2018mbn}.

\section{From Spain, with love}\label{sc:IAXOSpain}

The group of the University of Zaragoza, currently part of CAPA (Centro de Astropartículas y Física de Altas Energías of the University of Zaragoza), has a long trajectory in the field of the rare event searches, with deep expertise in low-background techniques and underground physics. The group started and developed the first underground facilities in the Spanish Pyrenees, which later gave birth to the Canfranc Underground Laboratory (Laboratorio Subterráneo de Canfranc, LSC), of which is a \emph{host group}. 

Zaragoza was a funding member of the CAST Collaboration. Among other tasks, it has been in charge of the operation of two detector systems. It is, at present, responsible for the operation and analysis of the IAXO pathfinder system in CAST, which is serving as the baseline design for the IAXO detectors\,\cite{Aznar:2015iia}.

CAPA is an active promoter of the IAXO project and is in charge of several working groups and work packages in the collaboration. The primary task of the group is the delivery of the baseline detector systems for IAXO. 
The group is a leader in the development of Micromegas readouts for their application in rare event searches. A leap forward in the relevant R\&D was the TREX-\emph{Novel concept in TPCs for Rare Events Searches in Astroparticle Experiments} project, financed with an ERC Starting Grant~\cite{Irastorza:2015dcb, Irastorza:2015geo}. One of the most important outcomes of TREX was the IAXO pathfinder detector, currently installed in CAST. The continuous improvements through a careful selection of screened materials, the use of an appropriate shielding, improving the fabrication of the Micromegas planes, the operation in combination with an x-ray focusing device, but also very importantly, the development of software tools for the optimization of background discrimination algorithms, resulted in the IAXO pathfinder registering the lowest background ever~\cite{Aznar:2015iia}.
The microbulk-read TPC for IAXO profit from all the experience gained with the IAXO pathfinder system. However, the background requirements for IAXO are strict, set to 10$^{-8}$\,c\,keV$^{-1}$\,cm$^{-2}$\,s$^{-1}$ which is more than one order of magnitude lower than the one of the IAXO pathfinder. To achieve that level, the IAXO baseline detectors will optimize the detector intrinsic radiopurity and minimize of the environmental background to the systems. 

Two of the important aspects towards this goal are the radioactivity measurements and the shielding of the detectors. The radiopurity working group of IAXO, led by CAPA, has an ongoing campaign of material screening for all detector groups, for which high purity Germanium Detectors that belong to CAPA and to the Radiopurity service of the LSC\,\cite{BANDAC2017127} are used. CAPA is also involved in the design of the passive shielding of the detectors and is participating in the development of the active shielding components.

Simulations of the different natural radiation components are used to support the hardware development of new and better designs and help on the background reduction and discrimination. As such, studies using the Geant4 simulation package\,\cite{Agostinelli:2002hh} have been performed to validate experimental detector measurements, and to optimize the configuration of passive and active shieldings installed at the detector setup. The group at the University of Zaragoza has developed a deep expertise in those activities, materializing it on a software ecosystem that allows to join those efforts and know-how into a single platform, REST-for-Physics.  This software is based on ROOT\,\cite{ROOT,ROOT2011} and it implements a core library, or framework, that defines appropriate interfaces oriented to process and analyze experimental or Monte Carlo event data in the field of the rare-event searches. Different REST-for-Physics libraries integrate appropriate tools and event processes for detector response, time signal processing, physical track reconstruction, axion ray tracing and other utilities that are used in a daily basis during the data processing and analysis of the results\,\cite{altenmueller2021restforphysics}. Developed within CAPA, it is a collaborative effort that has evolved to a generic framework to treat and store versioned data and simulation information. The expertise gathered on software development, simulations and analysis tools contributes to the IAXO physics program, with CAPA members coordinating the activities in the software and analysis groups.

In addition to the technical aspects, another area in which CAPA plays an important role is the theoretical motivation and make the most of IAXO's physics results. Part of the group, experts in fundamental physics and axion phenomenology, is contributing to the update and enhancement of the physics cases of BabyIAXO and IAXO. 

The current spokesperson of IAXO is a member of CAPA. 

Although CAPA has a big commitment with the Collaboration, it is not the only Spanish contributor to the project. 
In the detector systems, an innovative feature that is pursued within IAXO is the development of radiopure front-end electronics; a major contributor to this task is the Institut Ciènces del Cosmos of the Universitat de Barcelona (ICCUB). As part of its technology program, oriented to cosmology, astrophysics and particle physics, the ICCUB promotes experimental physcis and instrument development.

Another crucial aspect for helioscopes in general, is the system that will track the Sun with the required accuracy. The Centro de Estudios de Física del Cosmos en Aragón (CEFCA) is participating in the system aligning the magnet with the centre of the Sun. The CEFCA focuses on the technological development and operation of the Observatorio Astrofísico de Javalambre (OAJ) in Teruel, Spain, and on its scientific exploitation. 

The possibility of implementing in (Baby)IAXO microwave cavities to look for relic axions, in a similar way as in CAST with the Relic Axion Dark-Matter Exploratory Setup (RADES)~\cite{CAST:2020rlf}, is also being discussed (see Section~\ref{sc:observatory}). The RADES initiative is put forward by groups from the Universidad Politécnica de Cartagena, the Universidad de Valencia (and the Instituto de Física Corpuscular (IFIC)), the Centro Astron\'omico de Yebes, the LSC and CAPA, in addition to groups from CERN. More details are given in~\cite{RADESinSpain}.


\section{An Axion Observatory}\label{sc:observatory}

Next generation axion helioscopes represent an opportunity to extend the Standard Model of particle physics. The discovery of the axion would entail a major breakthrough on the tools we have for the understanding of the universe, representing an additional component on multi-messenger astronomy. 

IAXO might be seen as a \emph{swiss knife} for axion searches. Apart from the main research program, it represents a real infrastructure to explore new axion detection concepts and ideas. The versatility of this kind of experiments in the search of axions or ALPs has already been shown within the CAST experiment. The prospects of IAXO and its imminent construction has also motivated new ideas to exploit the IAXO physics program with a very promising future landscape on axion physics in case the axion is discovered.

If axions were discovered, they would constitute a potential candidate, together with neutrinos, to deeply study the physics of the Sun's interior.
The use of the Sun as an axion source transforms IAXO into an excellent tool to refine the well understood solar model. Recent studies highlight the potential role of axions (or ALPs) in solar physics. In particular, calculations of axion-to-photon conversions in the presence of macroscopic magnetic fields demonstrate that a non-negligible contribution to the axion flux is expected from the different solar magnetic regions due to longitudinal plasmons\,\cite{PhysRevD.101.123004} and transversal plasmons\,\cite{PhysRevD.102.123024}.
As it has been shown in a recent work\,\cite{PhysRevD.102.043019}, those production channels (shown in Figure\,\ref{fig:solar_flux}) might give access to measuring the unknown internal magnetic fields in the Sun. The identification of different axion atomic emissions from different isotopes would also allow to better determine the solar heavy isotope composition, or metallicity\,\cite{PhysRevD.100.123020}.

IAXO is conceived as an Observatory that would eventually expand its searches to other axion-related physics cases in a parasitic way. Looking for relic axions is one of them: installing a high-Q microwave cavity inside its magnet would turn IAXO into a haloscope. Some ideas towards such an implementation are being pursued~\cite{RADESinSpain}. 

Recently, the potential use of the IAXO setup for the detection of supernovae axions has been addressed\,\cite{Ge_2020}. Such a setup could be installed, for example, at the other end of the magnet (not used for solar tracking) and in the eventual case that an early supernovae warning would be received, the magnet could be redirected towards the supernovae explosion coordinates to catch the axion flux generated few hours after the supernovae collapse detection.

Finally, the role of helioscopes will be crucial in the field to unveil the axion properties. Most of the axion related processes allowed in the Sun's interior could be differentiated with a dedicated detection setup at IAXO, allowing to distinguish different solar components produced by different mechanisms (see Figure~\ref{fig:solar_flux}). The proper measurement of those components would lead to an independent measure of axion couplings to photons, electrons, or even nucleons\,\cite{PhysRevLett.75.3222,diluzio2021probing}, shedding light on the full theoretical description of the axion.

\paragraph{}
It is clear that axions and ALPs present attractive characteristics and properties and are drawing a lot of attention their way. IAXO will be prepared to discover them, looking in a good part of the parameter space; in parallel, other techniques plan to cover a complementary part of it. 

With an increasing part of the scientific community digging into axion physics, the odds are in our favor.

\end{paracol}


\vspace{6pt}



\authorcontributions{All authors have contributed equally to this article. All authors have read and agreed to the published version of the manuscript.}

\funding{ The Spanish groups mentioned in this paper were funded by the European Research Council (ERC) under the European Union’s Horizon 2020 research and innovation programme, grant agreement ERC-2017-AdG788781 (IAXO+), and from the Spanish Agencia Estatal de Investigaci\'on under grant FPA2016-76978-C3-1-P, the coordinated grant PID2019-108122GB, and the Maria de Maeztu grant CEX2019-000918of.}

\acknowledgments{
The authors would like to thank Maurizio Giannotti, visitor at University of Zaragoza in the framework of the Fulbright U.S. Scholar Program, for the fruitful discussions and comments on this review.}

\conflictsofinterest{The authors declare no conflict of interest.} 

\reftitle{References}


\externalbibliography{yes}
\bibliography{references}

%


\end{document}